\input psfig

\documentstyle{article}

\begin{document}

\vspace{1 cm}
\Large
\title{\bf Partitioning of a polymer chain between a confining cavity and a gel}

\vspace{1.8 cm}
\large
\author{Stefan Tsonchev$^{\dagger}$, Rob D. Coalson$^{\ddag}$, and Anthony Duncan$^{\S}$ \\ 
$^{\dagger}$Department of Chemistry, Northeastern Illinois University, Chicago, IL 60625 \\ $^{\ddag}$Department of Chemistry, University of Pittsburgh, Pittsburgh, PA 15260 \\ $^{\S}$ Department of Physics, University of Pittsburgh, Pittsburgh, PA 15260}
\date{}
\maketitle

\begin{abstract}
A lattice field theory approach to the statistical 
mechanics of charged polymers in electrolyte solutions [S.~Tsonchev, 
R.~D.~Coalson, and A.~Duncan, Phys. Rev. E {\bf{60}}, 4257, (1999)] is 
applied to the study of a polymer chain contained in a spherical cavity
but able to diffuse into a surrounding gel. The distribution of the polymer 
chain between the cavity and the gel is described by its partition coefficient,
which is computed as a function of the number of monomers in the chain, the 
monomer charge, and the ion concentrations in the solution.
\end{abstract}

\newpage
\section{Introduction}
The problem of partitioning of a polymer chain confined to move within large 
cavities embedded into a hydrogel has received attention with some
interesting experiments performed by Liu et al. \cite{Asher}. These authors 
investigated the so called ``entropic trapping'' phenomenon, which describes 
the preferential localization of a polymer chain within large cavities 
embedded in a hydrogel due to the larger conformational entropy of the chain 
in them. Therefore, this phenomenon has been suggested as a basis for 
potential 
new methods 
of polymer separation. In this context, the problem is also relevant and can 
lead to better 
understanding and possible improvement of many existing separation methods, 
such as membrane separation, filtration, gel electrophoresis, size exclusion 
chromatography, etc. \cite{Rod}, all of which utilize the dependence of 
macromolecular mobility through a network of random obstacles on molecular 
properties, such as molecular weight, monomer charge, electrolyte composition,
 etc. The practical 
importance of these techniques has motivated a number of investigations of 
polymer separation between cavities of different size [3--7].

In our earlier work \cite{Ts1,Ts2} we used lattice field theory calculations to
study polymer separation between two spheres of different size---a simplified 
model of the more complicated system of the polymer moving between large cavities embedded in a hydrogel, with the larger sphere playing the role of the cavities and the small sphere corresponding to the connecting channels in the gel. 
We investigated the dependence of the partition coefficient $K$, defined 
as the ratio of the average number of monomers in the two respective spheres, 
as a function of the total number of monomers in the chain, the excluded 
volume interaction between them, the monomer charge, and the concentration of 
electrolytes in the solution. Our results were qualitatively in accord with the experiments 
of Liu et al. \cite{Asher} and with related computer simulations \cite{Chern}.

In this work we apply the lattice field theory approach to the more complex 
system of a polymer chain moving within a large spherical cavity embedded in a 
network of random obstacles. 

In Section 2 of the paper, for continuity of the presentation, we review the 
lattice field theory of charged polymer chains in electrolyte solution 
\cite{Ts2}. In Section 3 we describe the Lanczos approach for finding the 
energy spectrum of the Schr\"odinger Hamiltonian problem (arising from the 
polymer part of the partition function \cite{Ts2}), and the resolvent 
approach for extracting the corresponding eigenvectors. Section 4 describes 
the numerical procedure for solving the mean field equations of the system, 
and in Section 5 we present and discuss our results. In Section 6 we conclude 
our presentation.

\newpage
\section{Review of Lattice Field Theory of Charged Polymer Chains in 
Electrolyte Solution}

In Ref. \cite{TCD} we derived
the following functional integral expression for the 
 full partition function of a charged polymer in an 
electrolyte solution with short-range monomer repulsion interactions 
\begin{equation}
 Z=\int 
D\chi(\vec{r})D\omega(\vec{r})e^{\frac{\beta\varepsilon}{8\pi}\int\chi\Delta\chi 
d\vec{r}-\frac{\lambda}{2}\int\omega(\vec{r})^{2}d\vec{r}+c_{+}\int 
e^{ie\beta\chi}d\vec{r}+c_{-}\int e^{-ie\beta\chi}d\vec{r}}Z_{Schr}(\chi,\omega) \, .
\label{Z}
\end{equation}
 Here,  $\beta{=}1/kT$ is the inverse temperature, $\varepsilon$ is the dielectric constant of the solution, $e$ is the proton charge, $\lambda$ is a 
measure of the strength of the excluded volume interaction, 
$\chi$ and $\omega$ are auxiliary fields, 
$c_{\pm}{=}e^{\beta\mu_{\pm}}/\lambda_{\pm}^{3}$ with $\mu_{\pm}$ and 
$\lambda_{\pm}$ being the chemical potentials and the thermal deBroglie 
wavelengths for the ions, respectively.  The polymer part  $Z_{Schr}(\chi,\omega)$ in
 (\ref{Z}) refers to a Euclidean-time ($T{=}M{=}$total number of monomers)
 amplitude for an equivalent Schr\"odinger
problem based on the Hamiltonian
\begin{equation}
  H\equiv -\frac{a_{p}^{2}}{6}\vec{\nabla}^{2}+\lambda\omega_{c}(\vec{r})
+\beta pe\chi_{c}(\vec{r}) \, ,
\label{eq:H}
\end{equation}
 where $a_p$ is the Kuhn length and $p$ is the charge per monomer.
The mean-field equations corresponding to  the purely-real saddle-point configuration fields 
$\chi_{c}=i\chi$, $\omega_{c}=i\omega$ are obtained by setting the variational derivative 
of the exponent in the full functional integral (\ref{Z}) to zero. For the case
 of a polymer with free ends (the only situation considered in this paper), the
 polymer amplitude $Z_{\rm Schr}$ can be written in terms of sums over
 eigenstates of $H$ as follows \cite{Ts2}:
\begin{eqnarray}
Z_{Schr}&=&{\int}dx_{i}dx_{f}\sum_{n}\Psi_{n}(x_{i})\Psi_{n}(x_{f})e^{-ME_{n}} \nonumber \\
&=&\sum_{n}A_{n}^{2}e^{-ME_{n}}\,{\equiv}\,e^{F_{pol}} \, ,
\label{eFpol}
\end{eqnarray}
where $E_{n}$ is the $n$-th energy eigenvalue,
\begin{equation}
A_{n}\equiv{\int}d\vec{r}\,\Psi_{n}(\vec{r}) \, ,
\label{An}
\end{equation}
and
\begin{equation}
F_{pol}=\ln\left(\sum_{n}A_{n}^{2}e^{-ME_{n}}\right) \, 
\label{Fpol}
\end{equation}
is the negative of the polymer contribution to the free energy.
Thus, the mean-field result for the negative of the total free energy is
\begin{equation}
  F=\int d\vec{r}\left\lbrace 
\frac{\beta\varepsilon}{8\pi}\left|\vec{\nabla}\chi_{c}\right|^{2}+\frac{\lambda}{2}
\omega_{c}^{2}+c_{+} e^{\beta e\chi_{c}}+c_{-} e^{-\beta e\chi_{c}}\right\rbrace
+F_{pol}(\chi_{c},\omega_{c}) \, .
\label{F}
\end{equation}
 Varying the functional (\ref{F}) with respect to the fields $\chi_c, \omega_c$ one obtains
 the mean-field equations
\begin{eqnarray}
  \frac{\varepsilon}{4\pi e}\vec{\nabla}^{2}\chi_{c}(\vec{r})&=&c_{+}e^{\beta 
e\chi_{c}(\vec{r})}
 - c_{-}e^{-\beta e\chi_{c}(\vec{r})}-p\rho(\vec{r})
\label{eq:pb} \, ,  \\
  \frac{a_{p}^{2}}{6}\vec{\nabla}^{2}\Psi_{n}(\vec{r})&=&\lambda 
\rho(\vec{r})\Psi_{n}(\vec{r})
+\beta pe\chi_{c}(\vec{r})\Psi_{n}(\vec{r})-(E_{n}-V_{m}(\vec{r}))\Psi_{n}(\vec{r}) \, ,
\label{eq:nlse}
\end{eqnarray}
where $\rho$, defined as
\begin{equation}
\rho(\vec{r}){\equiv}-\frac{\sum_{n,m}\frac{A_{n}\Psi_{n}A_{m}\Psi_{m}}{E_{n}-E_{m}}\left(e^{-ME_{n}}-e^{-ME_{m}}\right)}{\sum_{n}A_{n}^{2}e^{-ME_{n}}} \, ,
\label{rho}
\end{equation}
is the total monomer density.
The equations presented here apply
for polymer chains of arbitrary length, provided all (or a sufficient number) of states are
 included in the sums above. The single-particle potential $V_{m}(\vec{r})$ has been included to enforce an exclusion region 
for the monomers \cite{TCD}. Note that the 
parameters $c_{\pm}$ are exponentials of the 
chemical potentials $\mu_{\pm}$ for positively and negatively charged ions. 
The numbers of these ions must be fixed by suitably adjusting $c_{\pm}$  
to satisfy the relations
\begin{equation}
  n_{\pm}=c_{\pm}\frac{\partial \log{(Z)}}{\partial c_{\pm}}=c_{\pm}
\int e^{\pm\beta e\chi_{c}}d\vec{r} \, .
\end{equation}

  The advantage
of working with $F$ is that, as  shown in Ref. \cite{Ts2}, it has a 
unique minimum,  and thus, can be used to guide
a numerical search for the mean electrostatic and monomer
density fields.  Once the mean fields have been computed,
the defining relation ${\ln}Z{\cong}F(\chi_c,\omega_c)$ can
be used to obtain free energies of various types.
For example, the Helmholtz free energy $A$ (corresponding
to fixed numbers of monomers and impurity ions) is given by
\begin{equation}
\beta A = n_+ \ln c_+ + n_- \ln c_- - F(\chi_c,\omega_c) \, .
\end{equation}

Following the procedure of Ref. \cite{TCD}, we now move from the continuum to a
discrete 3-dimensional lattice by rescaling according to 
\begin{eqnarray*}
f(\vec{r})&{\rightarrow}&{\beta}e\chi_{c}(\vec{r})\\
\Psi_{N}(\vec{r})&{\rightarrow}&a_{l}^{3/2}\Psi_{N}(\vec{r})
\end{eqnarray*}
 and multiplying
Eq. (\ref{eq:pb}) by $a_{l}^{3}$ ($a_l$ being the lattice spacing). 
This leads to the following discretized version of equations 
(\ref{eq:pb}) and (\ref{eq:nlse}) on a 3D lattice: 
\begin{eqnarray}
\alpha\sum_{\vec{m}}\Delta_{\vec{n}\vec{m}}f_{\vec{m}}&=&\gamma_{+}e^{f_{\vec{n}}}-\gamma_{-}e^{-f_{\vec{n}}}-p\rho_{\vec{n}} \label{PBE} \, , \\
\frac{a_{p}^{2}}{6a_{l}^{2}}\sum_{\vec{m}}\Delta_{\vec{n}\vec{m}}\Psi_{N,\vec{m}}&=&\frac{{\lambda}M}{a_{l}^{3}}\rho_{\vec{n}}\Psi_{N,\vec{n}}+pf_{\vec{n}}\Psi_{N,\vec{n}}-E_{N}\Psi_{N,\vec{n}} \label{SE} \, ,
\end{eqnarray}
where
\begin{eqnarray}
\alpha&=&\frac{{\varepsilon}a_{l}}{4\pi{\beta}e^{2}} \, , \\
\gamma_{\pm}&=&\frac{n_{\pm}}{\sum_{\vec{n}}e^{\pm f_{\vec{n}}}} \, ,
\end{eqnarray}
and the wavefunctions are dimensionless and normalized according to
\begin{equation}
\sum_{\vec{n}}\Psi_{N,\vec{n}}^{2}=1 \; ;  \; 
\end{equation}
thus, the density $\rho_{\vec{n}}$ sums to the total number of 
monomers, $M$.

\newpage
\section{Extraction of Eigenspectrum and Eigenfunctions for Polymer Effective Hamiltonian}
  
   The simultaneous relaxation solution of Equations (\ref{PBE}) and (\ref{SE}) requires
 a rapid and efficient extraction of the eigenvalues and low-lying eigenvectors of the
 operator $H$, which amounts---once the problem has been set up on a discrete 
finite 3-dimensional lattice---to a large sparse real symmetric matrix. 

    We have found it convenient to use  distinct algorithms to extract the low-lying
 spectrum and eigenvectors of $H$ (typically we need on the order of 10--30 
 of the lowest states for the shortest polymer chains studied here, while for 
the longest polymer chains only one to three states suffice). The eigenvalues 
are extracted using the Lanczos technique \cite{lanc}.  Starting from a 
random initial
 vector $w_{0}\equiv v_{1}$, one generates a series of orthonormal vectors
 $v_1,v_2,...$ by the following recursion:
\begin{eqnarray*}
    v_{n+1} &=& w_{n}/\beta_{n} \, , \\
    n &\rightarrow& n+1 \, , \\
    \alpha_{n} &=& (v_{n},Hv_{n}) \, , \\
    w_{n} &=& (H-\alpha_{n}I)v_{n}-\beta_{n-1}v_{n-1}\, , \\ 
    \beta_{n} &=& \sqrt{(w_{n},w_{n})} \, ,
\end{eqnarray*}
where $\alpha_{n},\beta_{n}$ are real numbers, with $\beta_{0}=1$ and $v_{0}=0$. The
 matrix of $H$ in the basis spanned by $v_{n}$ is tridiagonal with the number $\alpha_{n}$
($\beta_{n}$) on the diagonal (respectively, super/sub diagonal). Carrying the Lanczos recursion
 to order $N$, diagonalization of the resulting $N{\times}N$ tridiagonal matrix leads, for large 
$N$, to increasingly accurate approximants to the exact eigenvalues of $H$. The
 presence of spurious eigenvalues (which must be removed by the sieve method
 of Cullum and Willoughby \cite{Cul}) means that typically a few hundred
 Lanczos steps must be performed to extract the lowest 30  or 40 eigenvalues of $H$
 (for dimensions of $H$ of order 10$^5$ as studied here) to double precision.

   Once the low-lying spectrum of $H$ has been extracted by the Lanczos 
procedure,
 as outlined above, the corresponding eigenvectors are best obtained by a resolvent
 procedure. Supposing $\lambda_{n}$ to be the exact $n$th eigenvalue of $H$ 
(as  obtained by the Lanczos method),  and $\psi_{n}$ the corresponding 
eigenvector,
 then for any random vector $\psi_{\rm ran}$ with nonzero overlap with $\psi_{n}$,
 the vector obtained by applying the resolvent
\begin{equation}
    \psi_{n,\rm approx} \equiv \frac{1}{\lambda_{n}+\epsilon-H}\psi_{\rm ran}
\end{equation}
is an increasingly accurate (unnormalized) approximant to the exact eigenvector
$\psi_{n}$ as the shift $\epsilon$ is taken to zero. A convenient algorithm for
performing the desired inverse is the biconjugate gradient method (see routine
linbcg in \cite{numrec}).  We have found the combination of Lanczos and
conjugate gradient techniques to be a rapid and efficient approach to the 
extraction of the needed low-lying spectrum.

\newpage
\section{Solving the Mean-Field Equations for a Polymer Chain Confined to Move 
within a Spherical Cavity Embedded in a Gel}

Equations (\ref{PBE}) and (\ref{SE}) are solved simultaneously using the 
following relaxation procedure \cite{TCD}. First, the Schr\"odinger
Eq. (\ref{SE}) is solved 
for $f_{\vec{n}}{=}0$ and ignoring the
nonlinear (monomer repulsion) potential term. 
The resulting $\Psi_{N,\vec{n}}$'s and corresponding energy levels
$E_N$ (wavefunctions and energy eigenvalues of a particle confined
to the cavity in a gel system) are used 
to calculate $\rho_{\vec{n}}$, then the Poisson-Boltzmann
Eq. (\ref{PBE}) is solved at each lattice 
point using a simple line minimization procedure \cite{walsh}.
The process is repeated and the 
coefficients $\gamma_{\pm}$ are updated after 
a few iterations until a predetermined accuracy is achieved. Then 
the resulting $f_{\vec{n}}$ is used in Eq. (\ref{SE}), which is solved
using the Lanczos method \cite{lanc}
for a new set of $\Psi_{N,\vec{n}}$'s to be used in calculating
an updated version of the monomer density
$\rho_{\vec{n}}$. 
This density is then inserted into Eq. (\ref{PBE}) and a new
version of $f_{\vec n}$ is computed.
For numerical stability, the updated $f_{\vec n}$ inserted into
Eq. (\ref{SE}) is obtained 
by adding a small fraction of the new $f_{\vec{n}}$ (just obtained
from Eq. (\ref{PBE})) to the old one
(saved
from the previous iteration). The same ``slow charging'' procedure is used for 
updating $\rho_{\vec{n}}$ in the nonlinear potential term of the Schr\"odinger 
equation (\ref{SE}).

This numerical procedure has been applied to the system of a polymer chain 
moving within a cavity embedded in a network of random obstacles. 
 We carve a spherical cavity of radius $10a_{l}$ in the middle 
 a cube with a side-length of $40a_{l}$ on $40^{3}$ lattice. The Kuhn length, 
$a_{p}=2a_{l}$, and in absolute units $a_{p}=5$\AA. Then the random obstacles are created by randomly selecting 20\% of the remaining lattice points in the cube outside the carved sphere to be off limits for the polymer chain. Thus,
the random obstacles take 20\% of the gel, 
that is, 80\% of the gel volume plus the cavity volume is available for the 
chain to move in. On the other hand, the impurity ions are 
free to move within the whole volume of the system. The monomer repulsion 
parameter $\lambda$ is fixed throughout the computations through the 
dimensionless parameter $\zeta$ by the following relation:
\begin{equation}
\zeta = 4\pi\frac{\lambda}{a_{p}^{3}} \, ,
\label{zeta}
\end{equation}
and the dimensionless parameter $\zeta=5$.

\newpage
\section{Numerical Results and Discussion}

We have computed the log of the partition coefficient 
$K\equiv\left<M_{1}\right>/\left<M_{2}\right>$, where $\left<M_{1}\right>$ 
and $\left<M_{2}\right>$ are the number of monomers in the spherical cavity 
and the remaining gel, respectively, as a function of the total number of 
monomers in the system, $M=\left<M_{1}\right>+\left<M_{2}\right>$, for 
varying monomer charge $p$ and varying number of ions in the system. In Fig. 1 we show the plot of $\ln{K}$ vs $M$ for two different monomer charges, $p=-0.1$ 
and $p=-0.2$ (all in units of $e$), and a fixed number of 600 negative impurity coions in the 
system, while 
the number of the positive counterions is fixed according to the condition for 
electroneutrality. 
\begin{figure}[!]
\psfig{file=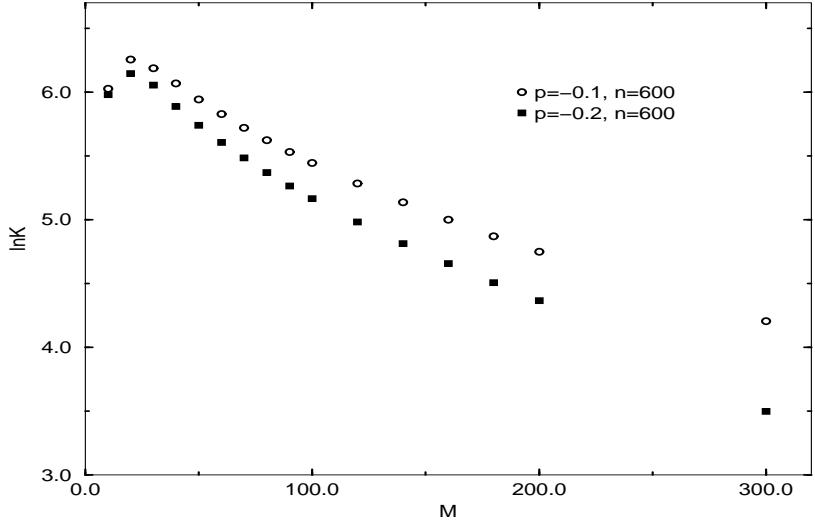,width=370pt,height=240pt,angle=270}
\vspace{-0.4 cm}
\caption{$\ln{K}$ vs $M$ for varying monomer charge $p$ and fixed number of 
negative impurity ions $n=600$, which corresponds to molar concentration $C\approx 0.996M$.}
\end{figure}
We see that the partition coefficient $K$ increases with $M$ for only the 
shortest polymer chains, goes through a turnover, and from then on decreases 
continuously as the number of monomers is increased. As in our previous work 
\cite{Ts1,Ts2}, 
we observe that smaller monomer charge leads to higher partition coefficient, 
due to the weaker repulsion between the monomers. In Fig. 2 we show how the 
partition coefficient varies as we vary the number of negative 
impurity ions in the system. As expected, the higher number of ions leads 
to better screening of the monomer charges, hence less repulsion and larger 
$K$.  
\begin{figure}[!]
\psfig{file=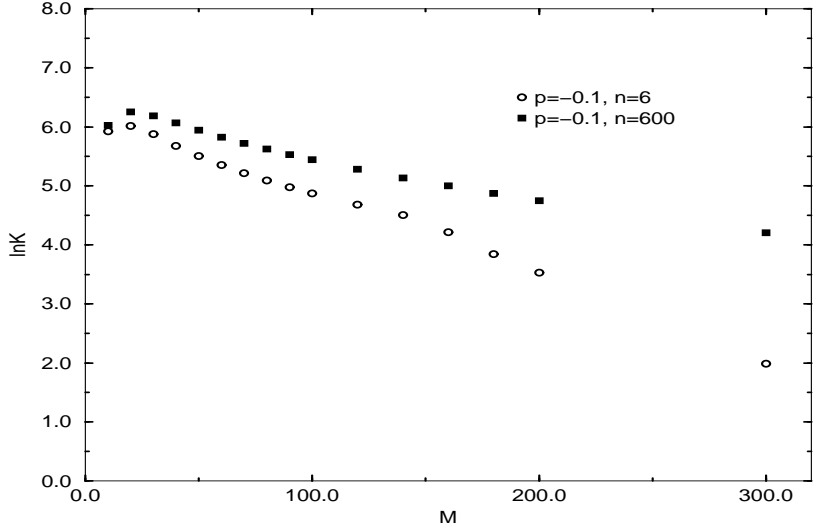,width=370pt,height=240pt,angle=270}
\vspace{-0.4 cm}
\caption{$\ln{K}$ vs $M$ for varying number of negative impurity ions $n$ 
and fixed monomer charge $p=-0.1$.}
\end{figure}
Qualitatively, this 
behavior is similar to what we observed in our previous work \cite{Ts1,Ts2}; 
however, the partition coefficient shown here decreases for almost the 
whole range of $M$, and is, in fact, much larger than the coefficients 
reported earlier \cite{Ts1,Ts2} for partitioning of a polymer chain between 
two spheres. This can be explained as a result of the much smaller  
voids that arise between the random obstacles in the gel outside of the 
spherical cavity, compared to the smaller of the two spheres treated in 
\cite{Ts1,Ts2},
which, in the Schr\"odinger language, means that, even though the volume 
available for the polymer chain outside of the spherical cavity is much 
greater than the volume of the cavity itself, the energy levels of the 
excited states which lead to a higher monomer density outside of the cavity 
are too high (due to the strong confinement in the narrow voids), so that the 
chain is largely confined to the cavity. Only 
for the cases of very large $M$ do we observe a non-negligible monomer 
density outside of the cavity. This is illustrated in Figs. 3 and 4, where 
we plot the averaged radial density of monomers starting from the center 
of the spherical cavity for the three different sets of monomer charge and impurity ion concentration parameters 
presented here. In Fig. 3 we plot the radial density for the case of 
relatively small number of monomers, $M=40$, and we see that virtually all 
of the monomers are confined to the spherical cavity, while in Fig. 4, which 
represents the case of $M=300$, we observe a small but non-negligible 
contribution to the monomer density from the region outside of the cavity. 
\begin{figure}[!]
\psfig{file=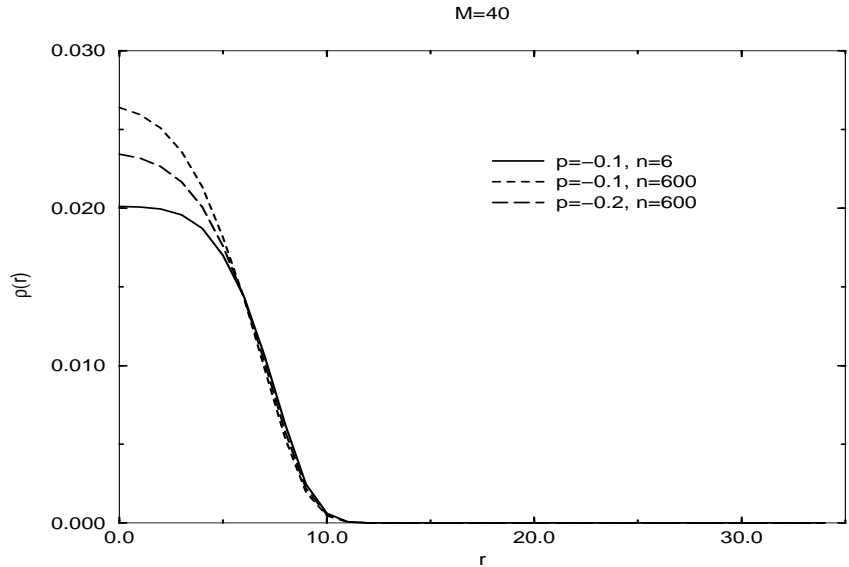,width=370pt,height=240pt,angle=270}
\vspace{-0.4 cm}
\caption{The average radial density $\rho(r)$ as a function of the distance 
from the center of the spherical cavity $r$ for the three sets of parameters 
considered here in the case of $M=40$. }
\end{figure}
\begin{figure}[!]
\psfig{file=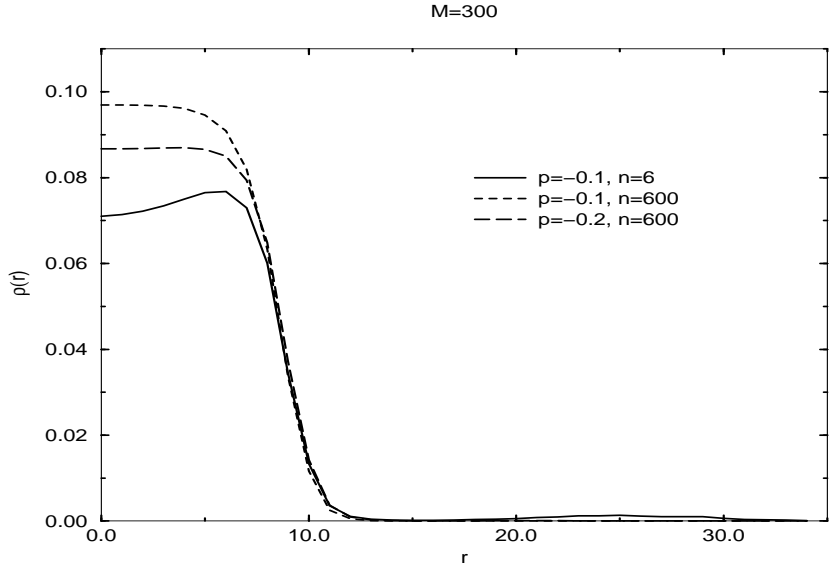,width=370pt,height=240pt,angle=270}
\vspace{-0.4 cm}
\caption{Same as in Fig. 3, but for $M=300$.}
\end{figure}
In Figs. 5 and 6 we show the plots of the electric potential $f(r)$ 
corresponding to the parameters of Figs. 3 and 4, respectively. We can 
qualitatively 
compare the results from Figs. 3--6 to our previous results in \cite{TCD}, 
where we computed the monomer density and the electric potential for a 
charged polymer chain confined to move within a sphere. It is clear that in 
both cases the shape of the monomer density distribution and the electric 
potential are quite similar, which is an illustration of the fact that 
the spherical cavity embedded in the gel does indeed act as an 
``entropic trap'' for the polymer chain, and for most of the range of 
reasonable physical parameters the system behaves approximately as a 
polymer chain in a spherical cavity. In Figs. 5 and 6 we see that, for the case of lower counterion numbers, the potential $f(r)$ drops to negative values at large radial distance. Nevertheless, it does approach (up to finite lattice size corrections) zero slope, or equivalently, zero electric field, consistent with the overall electrical neutrality of the system.
\begin{figure}[!]
\psfig{file=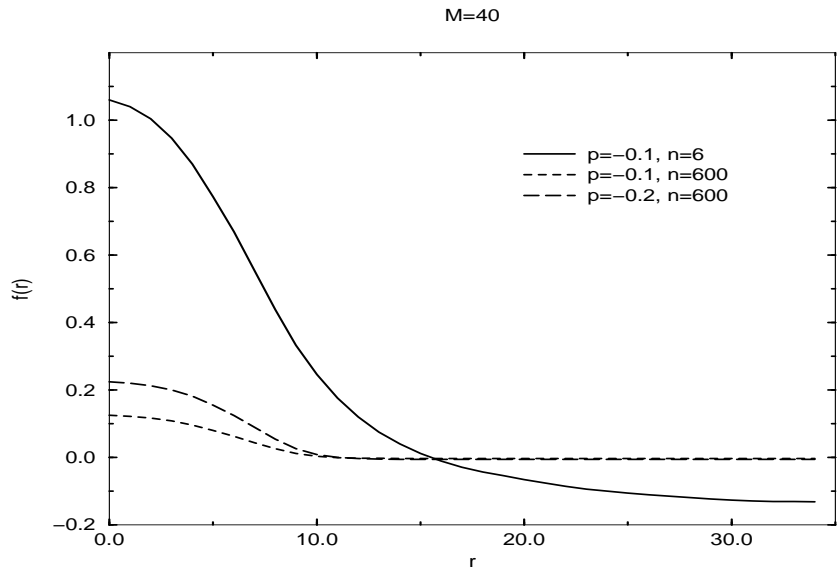,width=370pt,height=240pt,angle=270}
\vspace{-0.4 cm}
\caption{Electric potential $f(r)$ for the parameters corresponding to 
Fig. 3.}
\end{figure}
\begin{figure}[!]
\vspace{-0.5 cm}
\hspace{0.6 cm}
\psfig{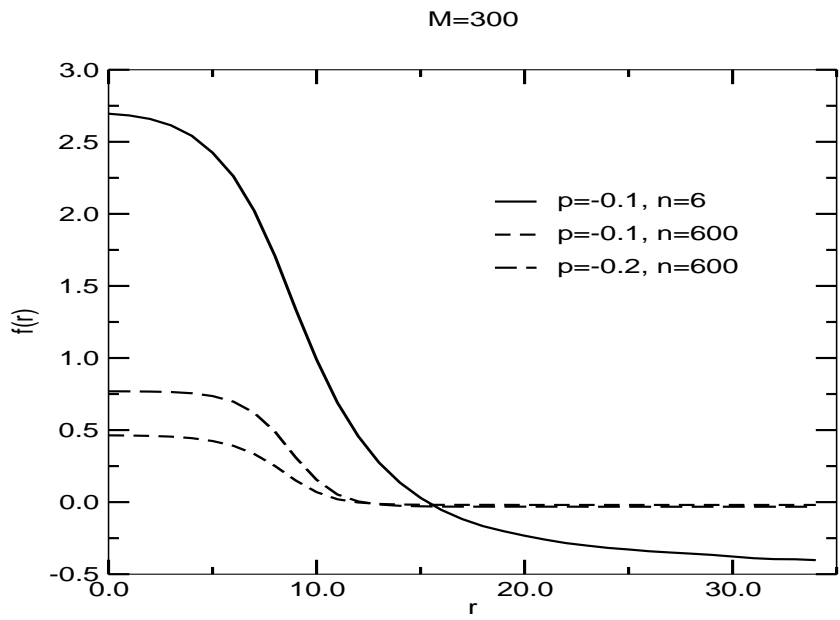}
\vspace{0.3 cm}
\caption{Electric potential $f(r)$ for the parameters corresponding to 
Fig. 4.}
\end{figure}

\newpage
\section{Conclusions}
We have applied a previously developed lattice field theory approach to 
the statistical mechanics of a charged polymer chain in electrolyte solution 
\cite{TCD,Ts1,Ts2} to the problem of a charged polymer chain moving in a 
spherical cavity embedded in a gel. This problem is more relevant to real 
experimental situations involving charged polymer chains in a complex 
environment than the two-sphere problem studied by us earlier \cite{Ts1,Ts2}. The results of this work demonstrate the capability of the 
approach to treat more complex systems of arbitrary shape in three dimensions,
and also confirm the expectations that a large spherical void carved out from 
a network of random obstacles can act as a ``trap'' for polymer chains, and 
therefore, may serve as a prototype for new methods of polymer 
separation based on macromolecular weight, monomer charge, and/or electrolyte 
composition. The results presented here confirm our previous contention 
\cite{TCD,Ts1,Ts2} that chains with smaller monomer charge would be easier to 
separate by a technique exploiting the idea of ``entropic 
trapping.'' Similarly, for chains with fixed monomer charge, a better 
separation would be achieved in solutions with higher impurity ion 
concentration---a parameter which is typically varied in the laboratory. 

It is important to note that the method used here is based on the mean field 
approximation, and therefore, the results should be considered only as 
qualitative. Nevertheless, one can expect that the long range of the 
electrostatic interaction and the strong confinement of the polymer chain 
inside the spherical cavity would result in weakly fluctuating density and electrostatic fields and 
would make the mean field approximation reliable \cite{Ts3}.

\vspace{0.5 in}
{\bf Acknowledgments}: R.D.C. gratefully acknowledges the support of NSF grant CHE-0518044. The research of A. Duncan is supported in part by NSF contract PHY-0554660.

\end{document}